\documentclass[twocolumn,superscriptaddress,showpacs,amsmath,amssymb,pra,longbibliography]{revtex4-1}

\usepackage[pdftex]{graphicx} 
\usepackage{dcolumn}  
\usepackage{bm}       
\usepackage[usenames,dvipsnames]{color}
\definecolor{URLCOL}{rgb}{0,0.52,0.83} 
\definecolor{LINKCOL}{rgb}{0.05,0.5,0} 
\definecolor{orange}{rgb}{0.6,0.3,0} 
\definecolor{CITECOL}{rgb}{0.25,0,0.48} 
\usepackage{epstopdf}

\usepackage[pdftex,bookmarks,breaklinks,bookmarksopen,bookmarksnumbered,colorlinks,linkcolor=LINKCOL,linktocpage,citecolor=CITECOL,urlcolor=URLCOL,pdfpagemode=UseOutline,pdftex]{hyperref}

\usepackage{fancyhdr}

\usepackage{booktabs}
\usepackage{natbib}

\definecolor{TITLECOL}{rgb}{0.1,0.2,0.7} 
\definecolor{SECOL}{rgb}{0.1,0.2,0.7} 
\definecolor{CONTENTSCOL}{rgb}{0.1,0.2,0.7} 
\definecolor{SSECOL}{rgb}{0.25,0,0.48} 
\definecolor{SSSECOL}{rgb}{0.2,0.08,0.53} 
\definecolor{FINCOL}{rgb}{0.01,0.3,0.07} 

\def\coloredtitle#1{\title{\textcolor{TITLECOL}{#1}}} 
\def\coloredauthor#1{\author{\textcolor{CITECOL}{#1}}} 

\definecolor{URLCOL}{rgb}{0,0.17,0.43} 
\definecolor{LINKCOL}{rgb}{0.05,0.4,0} 
\definecolor{CITECOL}{rgb}{0.35,0,0.48} 

\def\sss{\scriptscriptstyle\rm}

\def\bea{\begin{eqnarray}}
\def\eea{\end{eqnarray}}
\def\ben{\begin{equation}}
\def\een{\end{equation}}
\def\benu{\begin{enumerate}}
\def\enu{\end{enumerate}}

\def\bei{\begin{itemize}}
\def\eei{\end{itemize}}
\def\beit{\begin{itemize}}
\def\eit{\end{itemize}}
\def\benu{\begin{enumerate}}
\def\enu{\end{enumerate}}

\def\br{{\bf r}}

\def\x{_{\sss X}}
\def\c{_{\sss C}}
\def\s{_{\sss S}}

\def\F{_{\sss F}}
\def\xc{_{\sss XC}}

\def\H{_{\sss H}}
\def\ee{_{\rm ee}}

\def\dn{_\downarrow}

\def\n{n}
\def\t{^{\tau}}
\def\dv{\Delta v}
\def\dn{\Delta n}

\def\sec#1{\section{\textcolor{SECOL}{#1}}}

\def\dv{\Delta v}
\def\dn{\Delta\n}

\begin{document}

\coloredtitle{Exact conditions on the temperature dependence of density functionals}
\coloredauthor{K. Burke}
\affiliation{Department of Physics and Astronomy, University of California, Irvine, CA 92697}
\affiliation{Department of Chemistry, University of California, Irvine, CA 92697}
\coloredauthor{J. C. Smith}
\affiliation{Department of Physics and Astronomy, University of California, Irvine, CA 92697}
\coloredauthor{P. E. Grabowski}
\affiliation{Department of Physics and Astronomy, University of California, Irvine, CA 92697}
\coloredauthor{A. Pribram-Jones}
\affiliation{Lawrence Livermore National Laboratory, 7000 East Avenue, L-413, Livermore, California 94550}
\affiliation{Department of Chemistry, University of California, Berkeley, CA 94720}
\date{\today}

\begin{abstract}
Universal exact conditions guided the construction of most ground-state density
functional approximations in use today.  We derive 
the relation between the entropy and Mermin free energy density functionals 
for thermal density functional theory.  
Both the entropy and sum of kinetic and electron-electron repulsion
functionals are shown to be monotonically
increasing with temperature, while the Mermin functional is concave downwards.
Analogous relations are found for both exchange and correlation.
The importance of these conditions is
illustrated in two extremes: the Hubbard dimer and the uniform gas.
\end{abstract}

\maketitle
\def\D{F_{\sss I}}
\def\tp{^{\tau'}}
\def\tl{^{\tau,\lambda}}


\sec{Introduction}

Warm dense matter (WDM) is a rapidly growing multidisciplinary field that spans
many branches of physics, including for example astrophysics,
geophysics, attosecond physics, and nuclear physics\cite{KD09,CECO11,D13,MH13,RR14,SEJD14,SGLC15,KDBL15,C15}. 
In the last decade, quantum molecular dynamics, using DFT with electrons
at finite temperatures, has been extremely successful at
predicting material properties under extreme conditions, 
and has become a standard simulation tool in this field\cite{GDRT14}. 
Almost all such simulations use ground-state exchange-correlation (XC)
approximations, even when the electrons are significantly heated.
Thermal density functional theory (thDFT) was 
formalized by Mermin\cite{M65}, when he showed that the reasoning of Hohenberg
and Kohn\cite{HK64} could be extended to the grand canonical potential of
electrons coupled to a thermal bath at temperature $\tau$.  In recent times,
the Mermin-Kohn-Sham (MKS) equations of non-interacting electrons at finite
temperature, whose density matches that of the physical system, are being
solved to simulate warm dense matter\cite{KS65,PPGB14}.  In most of these calculations,
the ground-state approximation (GSA) is made, in which the exchange-correlation
(XC) free energy, which typically depends on $\tau$, is approximated by its
ground-state value.  Accurate results for the uniform gas are still being
found\cite{BCDC13,KSDT14,FFBM15,SGVB15}, which provide input to a thermal local density
approximation, but LDA is insufficiently accurate
for many modern applications, and thermal GGA's are being explored\cite{SD14}.

Many useful exact conditions
in ground-state DFT (relation between coupling constant
and scaling,
correlation scaling inequalities, exchange and kinetic scaling equalities, 
signs of energy components) were first derived\cite{LP85}
by studying the variational principle
in the form of the Levy constrained search\cite{L79}.  
Most of these conditions are satisfied (by construction) by the local
density approximation\cite{KS65} and have been used for decades to
constrain and/or improve more advanced approximations\cite{PBE96}.
Their finite temperature
analogs were derived in Ref. \cite{PPFS11} (see also
Ref. \cite{DT11}), and extended in Ref. \cite{PB15}.
Because the kinetic and entropic contributions always appear in the
same combination as the so-called kentropic energy (see Eq. (\ref{Axc}) and related text), such relations 
can never be used to extract either component individually.  

Many basic thermodynamic relations are proven via
quantum statistical mechanics\cite{Schwabl07}.   However, converting these to
conditions on density functionals is neither obvious nor trivial.
In the present work, we extend these methods to the dependence of the
Mermin functional (i.e., the universal part of the free-energy functional)
on the {\em temperature}, rather than
on the coupling constant or the scale of the density.  We find several new
equalities and inequalities which apply to
thDFT of all electronic systems.  This allows us to
separate entropic and kinetic contributions.  We show that 
the entropy density functional is monotonically increasing with temperature, as is the sum
of the kinetic and electron-electron repulsion density functionals, and that
the temperature derivative of the Mermin functional is the negative
of the entropy functional.  Thus the Mermin functional is concave downwards
as a function of
temperature.  Applying these conditions to the MKS system yields conditions
on the exchange-correlation free energy functionals.
Lastly, we illustrate all our findings in the two
extreme cases of the uniform gas and the Hubbard dimer.
We find a recent parametrization of the XC free energy of
the uniform gas violates our conditions, 
although only for densities that are so low as to be unlikely
to significantly affect any property calculated within thLDA.

\sec{Theory}

For a given average particle number, 
define the free energy of a statistical density-matrix $\Gamma$ as
\ben
A\t[\Gamma] = H[\Gamma] - \tau\, S[\Gamma],
\een
where $\hat H$ is the Hamiltonian operator, $S$ extracts the entropy, and
we use $\tau$ to denote temperature.
Define
\ben
\D[\Gamma]=T[\Gamma]+V\ee[\Gamma],
\een
where $\hat T$ is the kinetic energy operator and $\hat V\ee$ the electron-electron
repulsion operator.
Then
\ben
F\t[\Gamma] = \D[\Gamma] - \tau\,  S[\Gamma].
\een
The Mermin functional, written in terms of a constrained search, is\cite{PPFS11}
\ben
F\t[\n] = \min_{\Gamma\to\n} F\t[\Gamma],
\een
where the argument distinguishes functionals of the density from those of
the density-matrix.
The free energy of a given system can be found from
\ben
A\t = \min_\n \left\{ F\t[\n] + \int d^3r\, v(\br)\, \n(\br) \right\}.
\label{Atdef}
\een
We denote by $\Gamma\t[\n]$ the statistical density matrix that minimizes
$\hat F\t$ and yields density $\n(\br)$.
Then:
\ben
\frac{dF\t[\n]}{d\tau} = \frac{\partial F\t[\Gamma]}{\partial \tau}
+ \int d\Gamma\, \frac{\partial F\t[\Gamma]}{\partial \Gamma}\, 
\frac{d\Gamma\t[\n]}{d\tau},
\een
where all are evaluated at $\Gamma\t[\n]$.  
Because $\Gamma\t[\n]$ is the minimizer, its derivative with respect
to temperature (or any variable) vanishes.
Thus
\ben
\frac{dF\t[\n]}{d\tau} = - S\t[\n].
\label{SfromF}
\een
This is the DFT analog of the standard thermodynamic relation\cite{Schwabl07},
and implies
\ben
F\t[\n]=F^0[\n]-\int_0^\tau d\tau'\, S\tp[n],
\een
where $F^0[\n]$ is the ground-state functional\cite{HK64}. 
We note that Eq. (\ref{SfromF}) was derived in \cite{C15}, but only within lattice DFT.

Given a Mermin
functional (approximate or exact, interacting or not),
Eq. (\ref{SfromF}) defines what the corresponding entropy functional must be.
Since coordinate scaling\cite{PPFS11} can
separate the kentropic and potential contributions
in $F$, Eq. (\ref{SfromF}) allows the entropic and kinetic energy functionals to
be separated.   Alternatively, given an entropy functional,
Eq. (\ref{SfromF})
defines the temperature-dependence of the corresponding Mermin functional.
Since the entropy is always positive,
\ben
dF\t[\n]/d\tau \leq 0,
\label{dFineq}
\een
i.e., the Mermin functional is monotonically decreasing.

Now consider what happens when, for a given density and temperature
$\tau$, we evaluate the Mermin functional on the density matrix for that density
but at a different temperature.  By the variational principle,
Eq. (\ref{Atdef}),
\ben
F\t[\Gamma\tp[\n]] \geq F\t[\n],
\een
for any value of $\tau'$. Thus
\ben
\D [\Gamma\tp[\n]] - \tau\, S[\Gamma\tp[\n]] \geq \D\t[\n]  -\tau\, S\t[\n],
\een
or
\ben
\D\tp[\n] - \tau\, S\tp[\n] \geq \D\t[\n] -\tau\, S\t[\n].
\label{Dttp}
\een
Since this result is true for any pair of temperatures, we reverse $\tau$
and $\tau'$ to find:
\ben
\D\t[\n] - \tau'\, S\t[\n] \geq \D\tp[\n] -\tau'\, S\tp[\n].
\label{Dtpt}
\een
Addition of Eqs. (\ref{Dttp}) and (\ref{Dtpt}) yields
\ben
(\tau-\tau')\, (S\t[\n]-S\tp[\n]) \geq 0,
\een
so that the entropy monotonically increases with $\tau$:
\ben
{dS\t[\n]}/{d\tau} \geq 0.
\een
Combining this with Eq. (\ref{SfromF}) implies
\ben
{d^2F\t[\n]}/{d\tau^2} \leq 0.
\een
Thus $F\t[\n]$ is concave downwards.

We can also isolate the  behavior of $\D\t[\n]$.  If we multiply Eq. (\ref{Dttp})
by $\tau'$, and Eq. (\ref{Dtpt}) by $\tau$, and add them together, all entropic
contributions cancel, yielding
\ben
(\tau'-\tau)\, (\D\tp[\n]-\D\t[\n]) \geq 0,~~~~
{d\D\t[\n]}/{d\tau} \geq 0.
\een
Both $\D\t[\n]$ and
$S\t[\n]$ are monotonically increasing, but the net effect 
is that the Mermin free energy is decreasing.

Applying these conditions
to the Mermin-Kohn-Sham electrons\cite{PPGB14},
we find
\ben
{dF\s\t[\n]}/{d\tau} = - S\s\t[\n],
\label{SsfromFs}
\een
and the inequalities
\ben
\frac{dT\s\t[\n]}{d\tau}, 
\frac{dS\s\t[\n]}{d\tau}
\geq 0 \geq
\frac{dF\s\t[\n]}{d\tau}, 
\frac{d^2F\s\t[\n]}{d\tau^2}
\label{dFsineq}
\een
where subscript s denotes non-interacting, 
and $F\s\t[\n]=T\s\t[\n]-\tau\, S\s\t[\n]$.
Some of these relations have long been invoked for the uniform and
slowly-varying gases and for constructing orbital-free
density functionals (see Ref. \cite{VST12} and references therein),
but here they have been proven for every inhomogeneous system.

\sec{Illustration}

\begin{figure}[htb]
\includegraphics[width=\columnwidth]{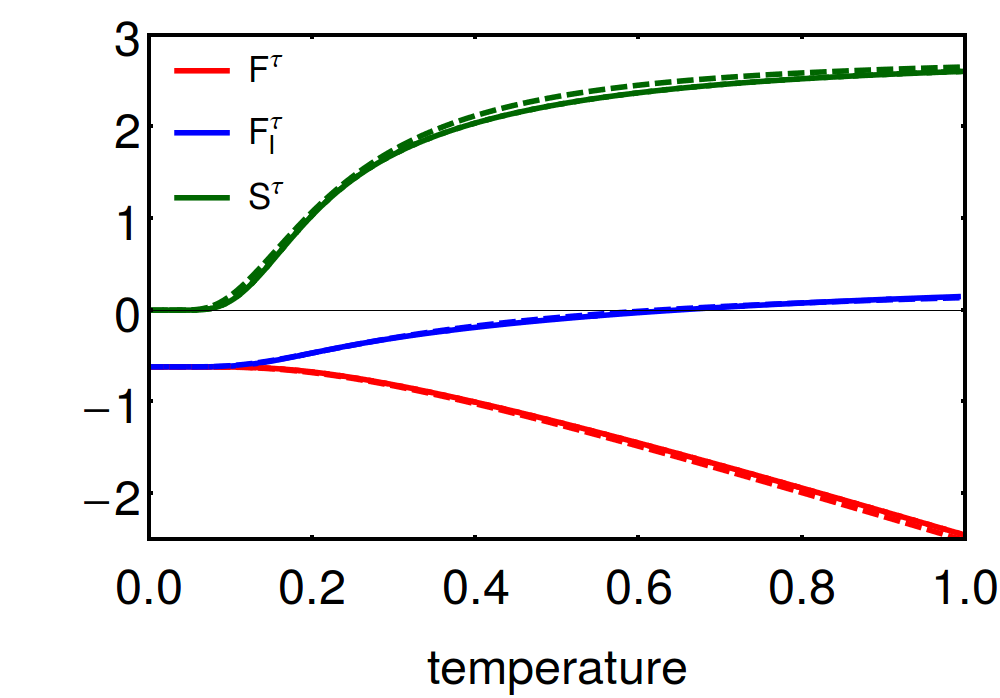}
\caption{Energy components for the Hubbard dimer in units of $2\,t$,
where $U=2\,t$ and $\dn=0$: $F\t,\D\t,S\t$, both interacting (solid)
and non-interacting (dashed).}
\label{Hub}
\end{figure}
To illustrate these results,
we calculate all energy components for an asymmetric Hubbard dimer, i.e. a two-site Hubbard model with a potential $v_1 = -v_2$,
as described in Ref. \cite{CFSB15} for the groundstate and \cite{SPB16} for the thermal system.  Here $t$ is the hopping parameter, $U$ the on-site
repulsion, and $\dn$ the difference in site occupations where the difference comes from having
an inhomogeneous potential $\dv = v_2 - v_1$.
This is the simplest possible model in which one can perform an exact thermal calculation,
including the exact thermal correlation components.
Fig. \ref{Hub}
shows the energy components, both interacting and non-interacting, as a function
of temperature for the homogeneous system with $\dn=0$.  All our exact conditions are satisfied for
many values of $\dn$ and $U$.

\begin{figure}[htb]
\includegraphics[width=\columnwidth]{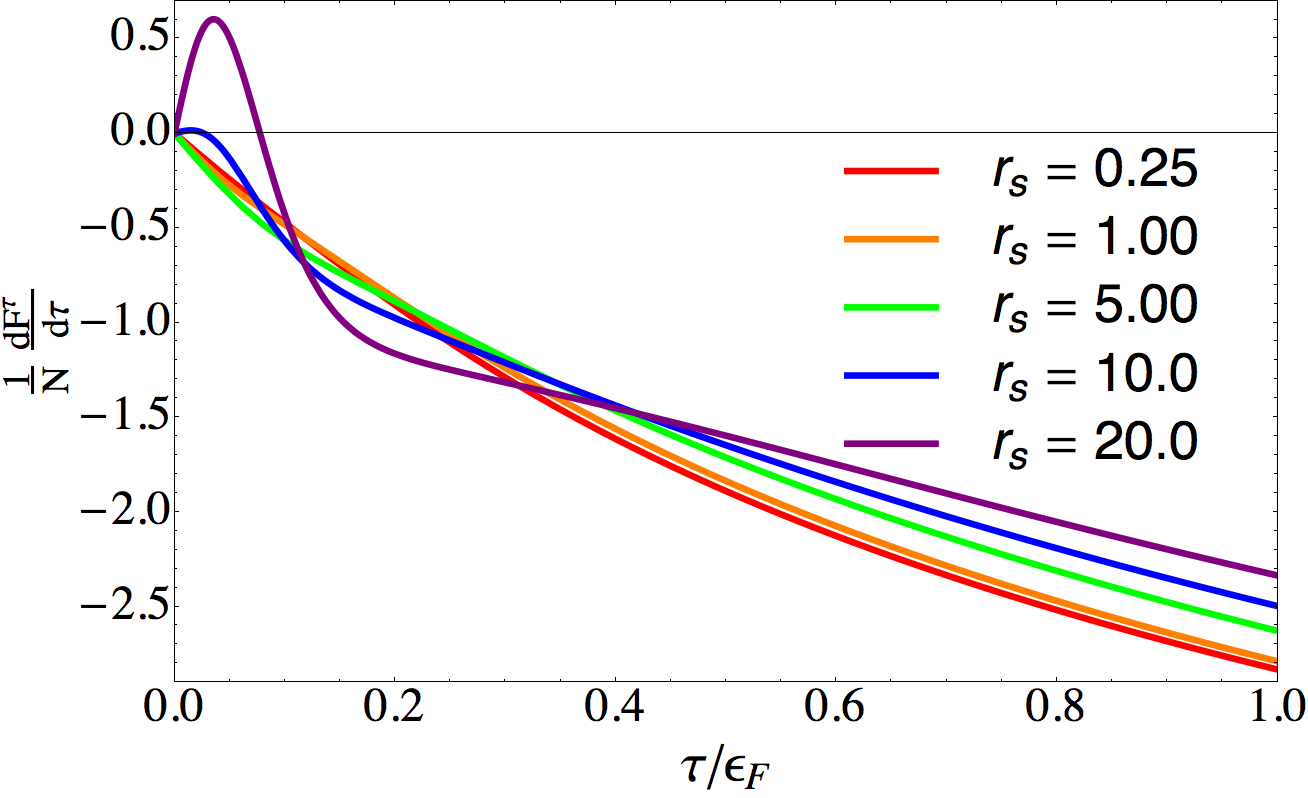}
\caption{Temperature dependence of the Mermin functional for spin-unpolarized uniform gas
for several values of the Wigner-Seitz radius $r\s$, using the XC
parametrization of Ref. \cite{KSDT14}, where $\epsilon\F$ is the Fermi
energy.}
\label{unifhi}
\end{figure}
At the other extreme is the uniform electron
gas and a modern parametrization of its free energy\cite{KSDT14}.
In the special case of a uniform density and potential, our formulas
become the same as the standard thermodynamic formulas.
In Fig. \ref{unifhi}, we plot the derivative of the
free energy per particle for fixed
density ($r\s$ value where $r\s = (3/(4\pi\n))^{1/3}$) as a function of temperature, on the scale of the
Fermi energy and in atomic units.
As $r\s \to 0$, these curves converge to
their well known\cite{DAC86} non-interacting value,
in which the derivative is negative and decreasing everywhere, in accordance with
Eq. (\ref{dFineq}).
Unfortunately, by decreasing the density so that XC effects become relatively
more important, we find that the parametrization violates our conditions for
$r\s > 10$.
Via Eq. (\ref{SfromF}), this implies that the entropy is
unphysically negative.
While such low densities are irrelevant to most practical
calculations using thLDA, parametrizations of
the uniform gas should build in
simple exact conditions such as ours.
Note that our restrictions apply only to continuous parametrizations.
The QMC data on which Ref. \cite{KSDT14} is based\cite{BCDC13}
is for the XC energy
at discrete values of the density, and so does not directly give the entropy.

For extremely high temperatures, sums
over KS eigenstates become impractical, and only pure DFT can be applied.
Because the uniform gas
satisfies our conditions, and because Thomas-Fermi (TF) theory uses local approximations to the
kinetic and entropic contributions which satisfy the conditions pointwise, we deduce
that TF theory satisfies our conditions.  
However, recent attempts to go beyond TF theory, such as using 
generalized gradient approximations for the energy\cite{KCST13,SD13,SD14}, should be tested 
for satisfaction of these constraints.

\sec{Exchange-Correlation}

In the final section of this paper, we apply this reasoning to the MKS method.
The Mermin functional is written in terms of the MKS quantities and a correction:
\ben
F\t[\n]=F\s\t[\n]+U\H[\n]+A\t\xc[\n],
\een
called the exchange-correlation (XC) free energy.  (The Hartree energy, $U\H[\n]$,
has
no explicit temperature dependence).  
The XC free energy is a sum of three components:
\ben
A\xc\t[\n] = K\xc\t[\n]+U\xc\t[\n]=T\xc\t[\n]-\tau S\xc\t[n]+U\xc\t[\n],
\label{Axc}
\een
where $U\xc\t$ is the potential contribution and $K\xc\t$ is the kentropic
contribution, which in turn consists of $T\xc\t$, the kinetic contribution,
and $-\tau S\xc\t$, where $S\xc\t$ is the entropic contribution.  

Subtract Eq. (\ref{SsfromFs})
from Eq. (\ref{SfromF}) to find
\ben
\frac{dA\xc\t[\n]}{d\tau} = - S\xc\t[\n],
\een
or
\ben
A\xc\t[\n]=E\xc[\n]-\int_0^\tau d\tau'\, S\tp\xc[\n].
\label{SxcfromAxc}
\een
All thermal XC effects are contained in the XC contribution to
the entropy.
This provides an intriguing alternative to the adiabatic connection
formula of Ref. \cite{PPFS11} or the thermal connection formula of Ref. \cite{PB15}.
Our inequalities
do not yield definite signs for XC quantities, just weak constraints
that would be difficult to impose universally on an XC approximation:
\ben
\frac{d T\xc\t}{d\tau} \geq -\frac{d T\s\t}{d\tau},~~~~
\frac{d S\xc\t}{d\tau} \geq -\frac{d S\s\t}{d\tau}.
\een
We can also combine these with the coupling-constant derivatives of
Ref. \cite{PB15} to find Maxwell-style relations:
\ben
\left( \frac{\partial U\xc}{\partial \tau} \right)_\lambda
= -\lambda\, \left( \frac{\partial S\xc}{\partial \lambda} \right)_\tau
\een
where $\lambda$ denotes evaluation at coupling-constant
$\lambda$, holding the density fixed\cite{PPFS11}.

Exchange can be
isolated by considering the limit of either weak interaction or scaling
to the high-density limit\cite{PPFS11}.  The exchange free energy is
\ben
A\x\t[\n]=V\ee[\Gamma\s\t[\n]]-U\H[\n]
\label{Axtdef}
\een
in a case of no degeneracies (the only case we consider here).  Because
$\Gamma\s\t$ minimizes the kentropy alone, to first order in
$\lambda$, kentropic corrections must be zero.  Thus
\ben
K\x\t[\n]=0,~~~T\x\t[\n]=\tau\, S\x\t[\n]\, 
=-\tau dA\xc\t[\n]/d\tau.
\label{Kxt}
\een
It may seem odd
to consider a kinetic contribution to exchange (impossible in the ground
state), but $T\x\t$ vanishes as $\tau\to 0$ in Eq. (\ref{Kxt}).  For
a uniform gas, the thermal exchange energy is well-known\cite{DAC86}.
But for our Hubbard dimer\cite{SPB16}, when $\langle N \rangle =2$, we
find $E\x[\n]=-U\H[\n]/2$, so that $T\x\t=S\x\t=0$.  

\begin{figure}[htb]
\includegraphics[width=\columnwidth]{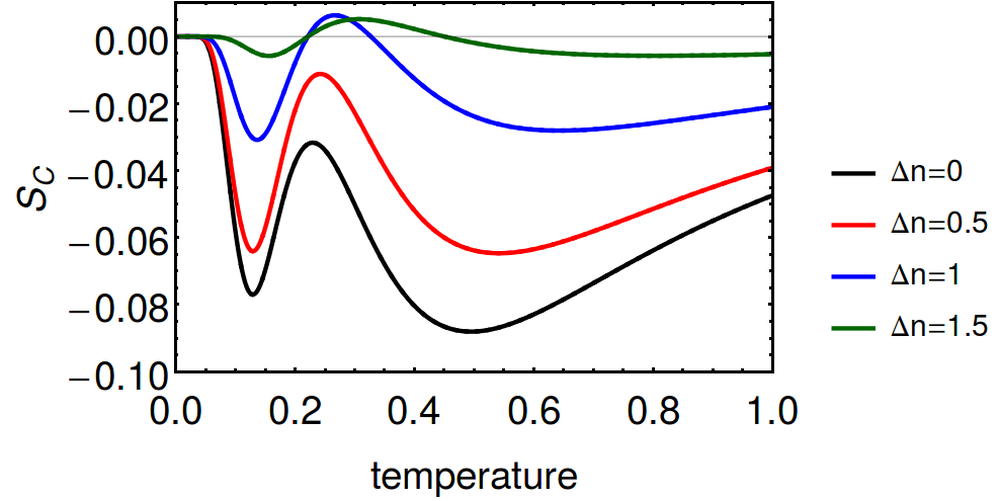}
\caption{Correlation entropy in the Hubbard dimer for several values of $\dn$ as a function
of temperature, in units of $2\,t$, where $U=2\,t$.}
\label{HubC}
\end{figure}
The results of Eq. (\ref{SxcfromAxc}) apply to correlation alone and
can be used in either direction, just as the relation for the full functional.
They are well-known for the uniform gas 
from statistical mechanics\cite{I82,PD84,PD00}.  But for an
inhomogeneous system, they are non-trivial, and so we illustrate them on 
the asymmetric Hubbard dimer.  In Fig. \ref{HubC}, we plot the entropic
correlation as a function
of temperature for several values of $\dn$, the occupation difference that arises from the asymmetric potential.
Eq. (\ref{SxcfromAxc}) is satisfied within numerical
precision.  The derivative of $S\c\t$ can change sign, even though
both $S\t(\dn)$ and $S\s\t(\dn)$ are monotonically increasing
(This explains the small dip seen in Fig. 7 of Ref. \cite{SPB16}).

\sec{Discussion and Conclusions}

Finally, we explain the apparent success of the
ground-state approximation (GSA) for $A\xc\t[\n]$
in MKS equilibrium calculations.  Almost
all present-day calculations of WDM use this approximation, and a recent
calculation on the Hubbard dimer\cite{SPB16} found that GSA
worked well when neither the temperature nor the strength of the
correlations were large (the conditions corresponding to most WDM calculations).
Now we explain why.  
Write
\ben
F^{\tau,{\rm GSA}}[\n]=F\s\t[\n]+U\H[\n]+E\xc[\n].
\een
Clearly, all temperature dependence is contained only in the KS part (usually
a very dominant piece).  Since the KS piece satisfies all the different 
inequalities and equalities, then so does any GSA calculation.  
But attempt to add corrections to a GSA calculation
by writing
\ben
A\xc^{\tau,{\rm GSA}}[\n]=E\xc^{\rm GSA}[\n]+ \Delta A\xc\t[\n].
\een
Only the thermal correction appears in the exact conditions we have derived,
since they all contain temperature derivatives.  But there is no simple
way to know
if the corrections will satisfy the exact conditions for all possible systems.
The only case would be using local approximations for all temperature-dependent
quantities, and then using energy densities from the uniform gas.  Thus a
TF calculation, with thermal LDA corrections, {\rm would} satisfy these
conditions, since they would be satisfied pointwise, as the uniform gas
satisfies these conditions for every density.  But in any MKS calculation
using approximate thermal XC corrections, this is not guaranteed.  
Unless special care is taken to guarantee satisfaction of our conditions,
{\em only} GSA automatically does this.  This is analogous to the situation in
TDDFT (at zero temperature): The adiabatic LDA, which ignores the history
dependence that is known to exist in the TDDFT functionals, satisfies most
exact conditions, while the time-dependent LDA (the Gross-Kohn approximation\cite{GK85})
violates several important constraints\cite{Db94}.  All this explains why the GSA
has been working well in many situations\cite{KD09,KDD08}.
The GSA appears to be correct in both the 
low- and high-temperature limits and, at least for model systems, reproduces 
the exact KS orbitals accurately\cite{SPB16}. Of course, this depends on the 
specific property being calculated and the acceptable level of error, and does not preclude 
moderate deviations, especially between these extremes, i.e., warm dense matter.
But any calculation that
includes, e.g., semilocal
thermal XC corrections, risks violating the exact conditions listed here that
GSA automatically satisfies, and should
be checked for such violations.
On the other hand, the Hartree-Fock approximation (or rather, the DFT equivalent,
called EXX\cite{KK08}), must satisfy the conditions since any expansion in powers of the
coupling constant up to some order must satisfy all our conditions.

To conclude, the formulas presented here are exact conditions applying to every
thermal electronic system when treated with DFT, and should guide the future
construction of approximate functionals.

\acknowledgments

The authors acknowledge support from the National Science Foundation (NSF)
under grant CHE - 1464795.
J.C.S. acknowledges support through the NSF Graduate Research fellowship program
under award \# DGE-1321846.
P.E.G. acknowledges support from the Department of Energy (DOE) under grant DE14-017426.
A.P.J's work was performed under the auspices of the U.S. Department of Energy by Lawrence Livermore National Laboratory under Contract DE-AC52-07NA27344.
A.P.J. was supported in part by the University of California President's Postdoctoral Fellowship.

\bibliography{Master,hubbard,thermal}
\label{page:end}
\end{document}